\documentclass[twocolumn,preprintnumbers,amsmath,amssymb]{revtex4}
\pdfoutput=1
\usepackage{graphicx}
\usepackage{dcolumn}
\usepackage{bm}

\begin{document}

\preprint{}

\title{Note on the dark matter explanation of the ARCADE excess and AMS data}

\author{Malcolm Fairbairn}
\email{malcolm.fairbairn@kcl.ac.uk}
\affiliation{Department of Physics, Kings College London}

\author{Philipp Grothaus}
\email{philipp.grothaus@kcl.ac.uk}
\affiliation{Department of Physics, Kings College London}

\date{\today}

\begin{abstract}
In this paper we show that positron data from AMS seems to rule out the explanation of the ARCADE isotropic radio background excess in terms of self-annihilating dark matter. In earlier works it was found that leptonic annihilation channels of light dark matter provide a good fit to the excess due to synchrotron emission of the final state particles. However, limits on the self-annihilation cross section derived from the positron data of AMS now severely constrain light self-annihilating dark matter and cross sections below that of a thermal relic are already tested for leptonic annihilation channels. Combining these two results, we conclude that an explanation of the excess in the radio background in terms of self-annihilating dark matter is excluded.
\end{abstract}

\maketitle
\paragraph*{Introduction}
The density of thermal relic dark matter (including weakly interacting massive particles) is set by its ability to annihilate with itself in the early Universe. The cross section required to obtain the correct relic density corresponds to new physics at the electroweak scale. For this reason it is hoped that this annihilation corresponds to the energy regime that is currently tested at the LHC and to cross sections that are searched for with current and next generation direct detection experiments.  A third place to search for such dark matter is in space, as the Standard Model annihilation products can give rise to observable signals across many wavelengths.

The balloon experiment ARCADE~2~\cite{Singal2009} has collected radio waves from the sky at frequencies between 3 and 90 GHz~\cite{Fixsen2009}. To extract a possible dark matter signal, a detailed model of the background has to be created and subtracted from the data. Background radiation comes from Galactic emission and also from extra-Galactic sources, such as the cosmic microwave background (CMB) or resolved point sources. Interestingly, after a careful background subtraction and including older maps at lower frequencies~\cite{Haslam1982,Reich1986,Roger1999,Guzman2011} in the analysis, an excess in terms of a power-law component remains, see e.g.~\cite{Fixsen2009,Fornengo2014}. The possible excess is the so-called ``ARCADE excess'', but note also~\cite{Subrahmanyan2013} for the claim of no excess in the data above background.

A possible explanation of the excess in terms of annihilating dark matter seems promising~\cite{Fornengo2011,Fornengo2012}. In earlier works it was found, that a good fit to the excess in isotropic radio waves can be achieved by light  dark matter annihilating into lepton final states and subsequent synchrotron emission of the final state particles. Surprisingly, a self-annihilation cross section close to that of the thermal WIMP paradigm is then needed~\cite{Fornengo2011,Hooper2012} making the dark matter explanation more plausible. In this paper, we confront this possible explanation further with limits derived from the AMS data~\cite{Aguilar2013, Bergstrom2013,Ibarra2013}, and show that the dark matter explanation is under great pressure and seems to be excluded. 

\paragraph*{The ARCADE Excess and Annihilating Dark Matter}
The observed spectrum from the ARCADE~2 Collaboration can be fitted well by the CMB plus an additional power-law spectrum:
\begin{equation}\label{eqn:temp}
T(\nu) = T_0 \,\frac{h\nu/(k\,T_0)}{\exp[h\nu/(k\,T_0)]-1}+T_s \left(\frac{\nu}{{\rm GHz}}\right)^{\alpha}~,
\end{equation}
with $T_0=2.729\pm0.004$~K as the temperature of the CMB. The ARCADE Collaboration published as best fits for the remaining parameters $\alpha=-2.62 \pm 0.04$ and $T_s=1.19\pm 0.14$~K~\cite{Fixsen2009}.

As was pointed out in~\cite{Fornengo2011}, and further investigated in~\cite{Hooper2012}, annihilating dark matter may explain this excess. The final state particles of dark matter self-annihilations are expected to emit synchrotron radiation that can possibly account for the additional observed power-law spectrum. In this paper we will not repeat the exercise of rederiving the dark matter mass and self-annihilation cross section necessary for a given annihilation channel to fit the excess observed by ARCADE. Instead, we take the robust results from reference~\cite{Hooper2012} and give a short summary of the analysis performed and the results.

The spectrum of synchrotron radiation from dark matter annihilations is given by:
\begin{eqnarray}\label{eqn:syn_em}
\frac{d\phi_{\rm syn}}{dE_{\rm syn}} &=& \frac{\sigma v}{8 \pi}\frac{c}{H_0} \frac{\bar{\rho}^2_{\rm DM}}{m^2_{\rm DM}} \int dz (1+z)^3 \frac{\Delta^2(z)}{h(z)}\frac{dN_{\rm syn}}{dE_{\rm syn}}\nonumber\\
&\times&E_{{\rm syn},0}(1+z)~,
\end{eqnarray}
where $dN_{\rm syn}/dE_{\rm syn}$ is the spectrum of synchrotron emission per dark matter annihilation, $z$ the redshift, $\sigma v$ the annihilation cross section, $H_0$ today's Hubble constant, $m_{\rm DM}$ the dark matter mass, $\bar \rho_{\rm DM}$ the averaged cosmological dark matter density, $h(z)=\sqrt{\Omega_{\Lambda}+\Omega_M (1+z)^3}$ and $\Delta(z)$ is the averaged squared overdensity of dark matter that depends on the halo mass function, on the dark matter density profile and on substructures within the halos.  The quantity $dN_{\rm syn}/dE_{\rm syn}$ depends on the injected electron and positron spectrum and through the energy losses also on the magnetic field in the environment of injection. To obtain the steady-state spectrum for $dN_{\rm syn}/dE_{\rm syn}$, the diffusion-loss equation has to be solved. This differential equation is:
\begin{eqnarray}
0&=&\vec{\nabla}\cdot [K(\vec{x},E_e) \vec{\nabla} \frac{dN_e}{dE_e}(\vec{x},E_e)]\nonumber \\ 
&+&\frac{\partial}{\partial E_e} [b(\vec{x},E_e) \frac{dN_e}{dE_e}(\vec{x},E_e)]+Q(\vec{x},E_e)~.
\end{eqnarray}
Here we have denoted $b(\vec{x},E_e)$ as the energy-loss rate which is mainly dominated by inverse Compton scattering and synchrotron emission and $K(\vec{x}, E_e)$ as the diffusion parameter. $Q(\vec{x}, E_e)$ is the source term and for annihilating dark matter given by $Q=\sigma v /(2 m_{\rm DM}^2) \rho_{\rm DM}^2 dN_{e,\rm Inj}/dE_e$. The injection spectrum of annihilating dark matter, $dN_{e,\rm Inj}/dE_e$, may be found in~\cite{Cirelli2010}. If diffusion can be neglected (e.g. for electrons and positrons that lose energy quickly in galactic magnetic fields), the steady-state solution of the electron spectrum is
\begin{equation}
\frac{dN_e}{dE_e}(E_e) = \frac{\sigma v \rho_{\rm DM}^2}{2 m^2_{\rm DM} b(E_e)} \int^{\infty}_{E_e} dE'_e \frac{dN_{e,{\rm Inj}}}{dE}(E'_e)~.
\end{equation}
The synchrotron emission from dark matter annihilation is obtained by combining equations~\ref{eqn:syn_em} and the steady-state electron spectrum. For detailed information about the halo mass functions and treatment of halo substructures we refer the reader directly to reference~\cite{Hooper2012}. Here, we give a short discussion about the applied parametrization of magnetic fields. 

In general, intracluster magnetic fields are poorly understood and introduce uncertainties when estimating the dark matter parameters that can give a good fit to the ARCADE data. The presence of these magnetic fields has been confirmed by the measurement of diffuse synchrotron emission from galaxy clusters. While their origin is still debated, conclusions about their strengths can already be drawn from different approaches. See, for example, references~\cite{Govoni2004,Ferrari2008} for summaries.

The ratio of the synchrotron flux and Compton produced X-ray flux allows estimations of the averaged magnetic field in galaxy clusters. Using this method, values of about 0.1 to 0.3 $\mu$G have been deduced. Also, rotation of polarized emission of radio sources within or behind a cluster opens up the possibility to constrain the averaged magnetic field strength along the line of sight due to Faraday rotation. This method points to magnetic field values of a few $\mu$G. Larger magnetic fields (tens of $\mu$G) have been found in the high density cooling-core regions.

Note here, that the values found using the Faraday effect are much larger compared to the ones deduced from the ratio of synchrotron to X-ray fluxes. This is not a contradiction, however, because no averaging over the large cluster volume is performed in the second method. Whereas it is still true that the intracluster magnetic fields are not well understood yet and more data is needed to be more confident about their strength, reasonable estimations based on both discussed methods are possible. The explanation of the ARCADE excess in terms of annihilating dark matter surely depends on the assumptions on these magnetic field strengths such that uncertainties on the resulting good parameter regions are unavoidable.

Reference~\cite{Fornengo2011} assumed a magnetic field which is constant throughout space and time with a value of 10 $\mu$G, which lies in the region expected from observations. A more evolved parametrization of the magnetic fields depending on the size of the halo and an added strong increase of the field values towards the halo centers has been adapted in reference~\cite{Hooper2012}. Additionally, the authors have performed a marginalization over the magnetic field strengths that significantly widen the good fit parameter regions and hence respect the uncertainties introduced from the unknown magnetic field strengths.
 
As both works assume magnetic field strengths consistent with the given experimental indications and point towards similar dark matter parameter regions, the derived values can be taken to be robust and confronting them with further data, in this paper coming from AMS, is reasonable and also necessary.

In the future, however, we could learn that the intracluster magnetic fields are much stronger than assumed in these works, lying beyond the range taken for the marginalization. Such stronger fields would enhance synchrotron emission and allow for smaller annihilation cross sections of dark matter as considered in this paper. However, at the moment, there are no such indications of much stronger magnetic fields.

Another source of uncertainty are substructures in halos. Clumped regions can boost synchrotron emission and a wrong treatment could overestimate the necessary annihilation cross section to fit the ARCADE data. However, as substructures are in the outer regions of halos and get disrupted when they fall into the center, their surrounding magnetic fields are weak and the impact of the clumping factor on synchrotron emission is expected to be moderate. From reference~\cite{Hooper2012} we take the set of parameter regions that were derived with the most optimistic boost factor, resulting in the smallest possible annihilation cross section that is most difficult to constrain.

Earlier works found that annihilations into lepton final states can give a good fit to the radio excess, whereas solutions in terms of $b \bar{b}$ give a too soft spectrum and are therefore less appealing as a solution of the ARCADE anomaly ~\cite{Fornengo2011,Hooper2012}. Reference~\cite{Hooper2012} confronted the annihilating dark matter explanation further with constraints from gamma-ray data~\cite{fermi_gamma,fermi_gamma2} and observations of dwarf spheroidals~\cite{fermi_dwarf,Geringer2011} excluding dark matter masses above 50~GeV because a too large gamma-ray flux would be present, independent of the annihilation channel. Also, it was shown that a pure annihilation into tau-antitau pairs is excluded by the gamma-ray data already.

Recently, AMS data has been reinvestigated in terms of annihilating dark matter by performing a spectral analysis~\cite{Bergstrom2013,Ibarra2013}. Strong constraints are derived on self-annihilating dark matter with lepton final states, especially for electron-positron pairs. For dark matter masses below 50~GeV the thermal annihilation cross section of the WIMP paradigm ($\sigma v = 3\times 10^{-26}$cm$^3$ s$^{-1}$) is excluded by more than an order of magnitude for dark matter candidates which annihilate into electron-positron or muon-antimuon final state particles. Annihilations into tau-antitau pairs are also under pressure, but within the quoted uncertainties the WIMP thermal cross section is still compatible with observations. In this work we will use these limits and check their compatibility with the dark matter explanation of the ARCADE excess.
\begin{figure}
\includegraphics[height=6.5cm]{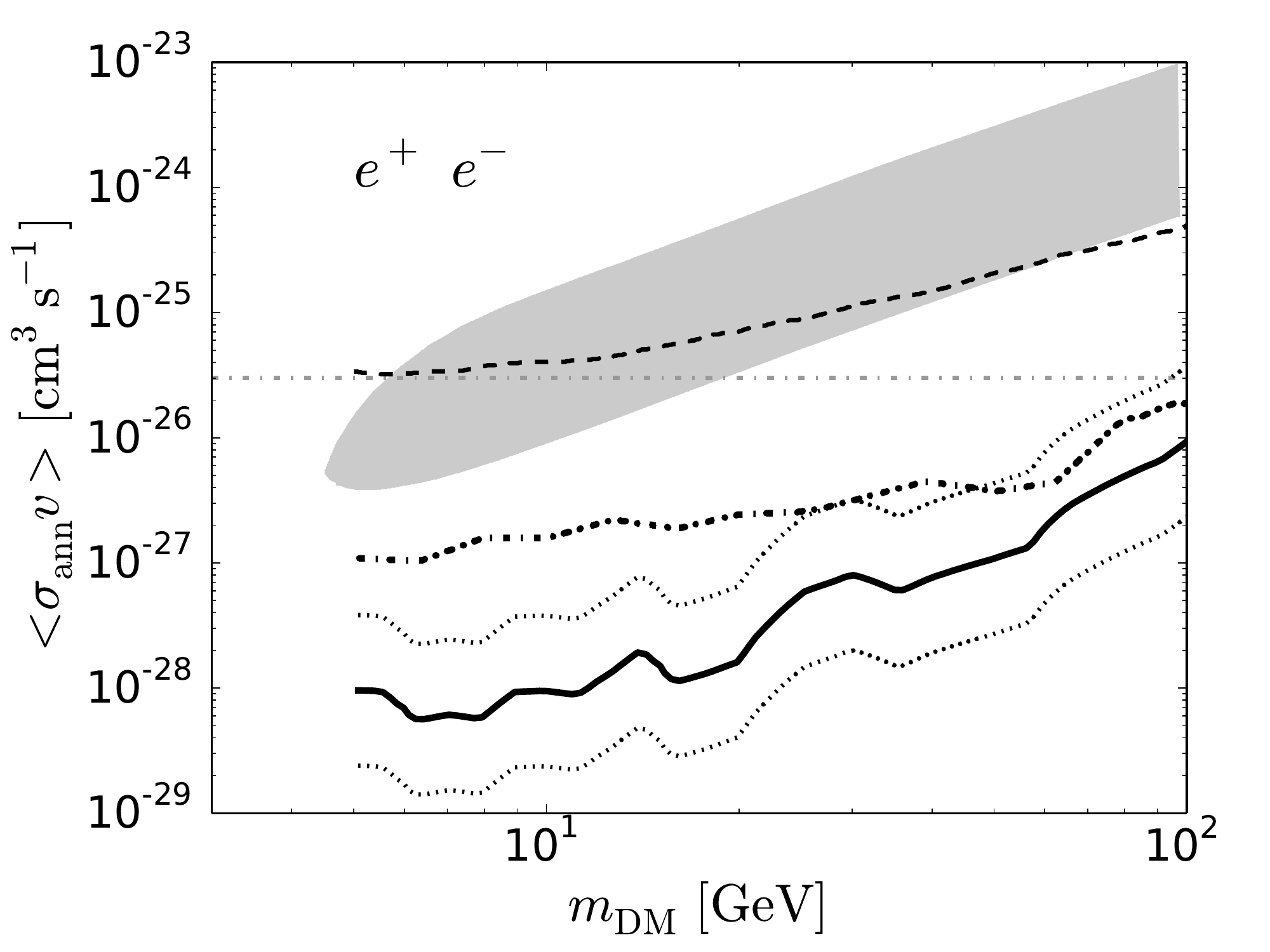}
\caption{\label{fig:electron} Dark matter mass and self-annihilation cross section that can explain the ARCADE excess for annihilation into electron/positron pairs (gray shaded region)~\cite{Hooper2012}, limits on the self-annihilation cross section from dwarf spheroidal galaxies (black-dashed line)~\cite{fermi_dwarf,Geringer2011} plus limits derived from the AMS positron fraction (black-solid and black-dotted lines)~\cite{Bergstrom2013} and positron flux (black-dash dotted line)~\cite{Ibarra2013}. It is apparent that the dark matter explanation of the excess is ruled out.}
\end{figure}

Uncertainties when analyzing the AMS data come from the unknown local dark matter density and also from the energy loss of electrons or positrons in the local magnetic fields. To account for these uncertainties, a range for the local dark matter density between 0.25 and 0.7 GeV cm$^{-3}$ and a range for the sum of the local radiation and magnetic field energy densities between $U_{\rm rad} + U_{\rm B} = (1.2 - 2.6)$ eV cm$^{-3}$ has been considered in~\cite{Bergstrom2013}. This then results into a band of exclusion limits on the annihilation cross section rather than a single line. We respect these uncertainties and show the exclusion bands.

In the Supplemented Material of ~\cite{SupplBergstrom2013}, the authors further discuss the impact of a nonsmooth background that might arise from a collection of pulsars and would make the search for a dark matter peak in the spectrum much more difficult. Indeed, they find that limits may weaken by up to a factor of 3, but only for the case of electro-positron final states. In the case of muon final states, the dark matter peak in the AMS data is expected to be much broader than a collection of pulsar peaks, such that the signal is sufficiently different and easily distinguished from the background. There was no indication of any peaked structure in the AMS data. Solar modulation may affect the diffusion of the low-energetic part of the positron flux and limits become less certain below a dark matter mass of 5~GeV and are not presented in~\cite{Bergstrom2013}.

The AMS data has independently been used in~\cite{Ibarra2013} to set limits on the dark matter annihilation cross section by using only the positron flux in order to reduce the dependence on the electron flux. These limits are significantly weaker than the ones coming from the positron fraction, but still strong enough to test thermal WIMP cross sections for light dark matter. When considering also the positron fraction, the authors find comparable limits to~\cite{Bergstrom2013}.
\begin{figure}
\includegraphics[height=6.5cm]{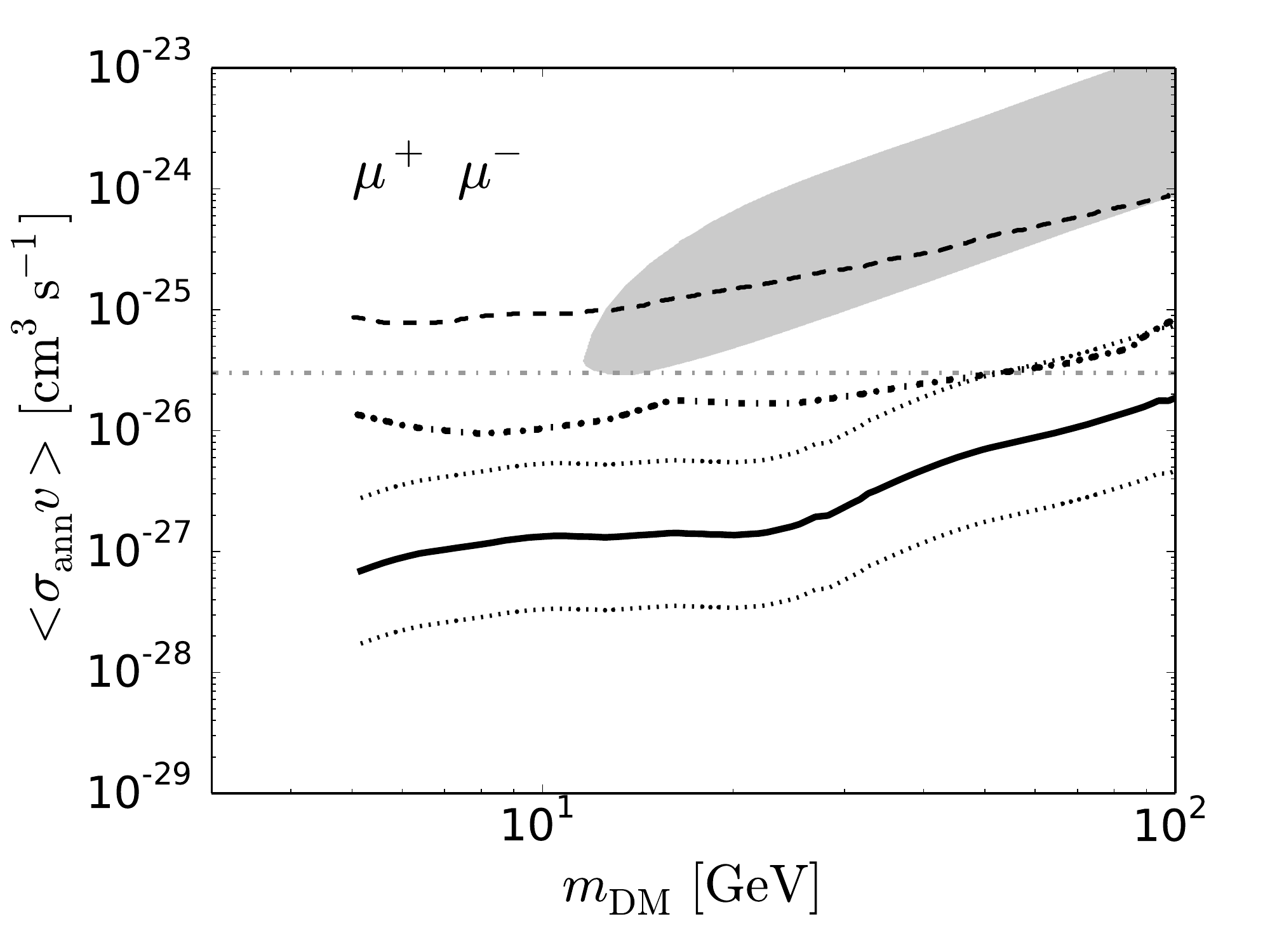}
\caption{\label{fig:muon} The same as figure~\ref{fig:electron} but with annihilation into muon/anti-muon pairs.}
\end{figure}

\begin{figure}
\includegraphics[height=6.5cm]{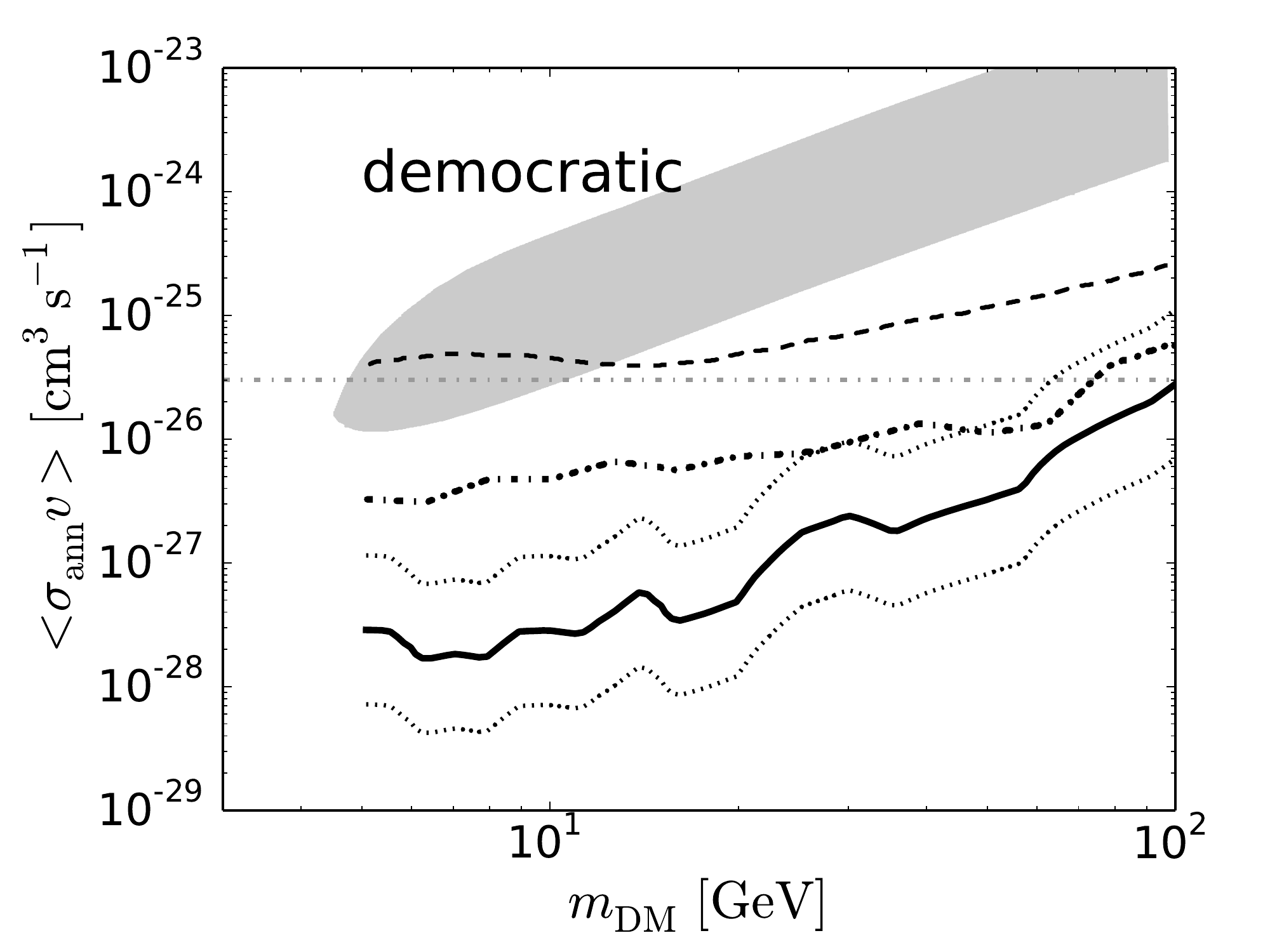}
\caption{\label{fig:democratic} The same as figure~\ref{fig:electron} but with dark matter annihilating in equal parts into electrons, muons and taus.}
\end{figure}

\paragraph*{Results} In figures~\ref{fig:electron},~\ref{fig:muon} and~\ref{fig:democratic} we present the regions of parameter space that can give a fit of the ARCADE excess as gray shaded regions and limits on the self-annihilation cross section of dark matter from dwarf spheroidal data as a black dashed line, following reference~\cite{Hooper2012}. Also, we show exclusion limits coming from the AMS positron fraction from reference~\cite{Bergstrom2013} as a black solid line and indicate the uncertainties quoted in this reference as black-dotted lines around the central exclusion value. Additionally, we draw the exclusion limits arising from the positron flux alone as a dash dotted black line~\cite{Ibarra2013}. The annihilation cross section of a standard WIMP to produce the correct relic abundance is indicated as a gray horizontal line.

For pure annihilation into electron-positron pairs or muon-antimuon pairs, the good region for the ARCADE excess is clearly excluded by the positron data of AMS. Both regions are almost an order of magnitude above even the most conservative exclusion limit, see figure~\ref{fig:electron} and~\ref{fig:muon}, respectively. For the case of electron-positron final states, a small window of less than a GeV is left open because limits from~\cite{Bergstrom2013,Ibarra2013} do not apply in this small dark matter mass region. We point out again that a nonsmooth background from pulsars would indeed weaken the limits on electron-positron final states by a factor of about 3. However, even these weakened limits would still put the ARCADE explanation into great tension with observations.

The same conclusion can be drawn for the democratic case, see figure~\ref{fig:democratic}, in which annihilation proceeds into all three lepton flavors with equal probability. For this scenario we simply rescaled the limits on the annihilation into electron-positron pairs by a factor of 3 to account for the different branching ratio. Note again that a pure annihilation into tau-antitau pairs is already excluded by gamma-ray data and that $b \bar{b}$ final states provide a much worse fit to the ARCADE data than lepton final states and are therefore less appealing~\cite{Fornengo2011,Hooper2012}.

Earlier works had shown that the isotropy of the excess is an indication against a dark matter explanation~\cite{Holder2012} and that the smooth intergalactic distribution of dark matter requires too large magnetic fields to explain the excess via annihilations~\cite{Cline2012}. 

We see that all possible scenarios that can explain the excess observed in the radio wave background in terms of self-annihilating dark matter are inconsistent with positron data from AMS and other explanations are now necessary.

\paragraph*{Conclusions} In this paper we assumed that the excess that is found in isotropic radio data is due to self-annihilating dark matter whose annihilation products emit synchrotron radiation and fit the observed power-law excess above the expected background. As was shown in earlier works~\cite{Fornengo2011,Hooper2012} only leptonic annihilation channels can provide a good fit to the data and are already under pressure from cosmic ray data. We confronted this scenario further with limits on the dark matter self-annihilation cross section derived from the positron data of AMS~\cite{Bergstrom2013,Ibarra2013} that heavily constrain light dark matter with leptonic annihilation channels. We presented in figure~\ref{fig:electron},~\ref{fig:muon} and~\ref{fig:democratic} that the explanation in terms of annihilating dark matter is ruled out by about an order of magnitude. Hence, other explanations have to be found.

We remarked that intracluster magnetic fields could turn out to be much stronger than assumed in the analyses such that energy loss of annihilation products into synchrotron radiation would be increased and smaller annihilation cross sections, that might fall into the allowed region of parameter space, would be allowed.

\begin{acknowledgments}
We thank Dan Hooper and Torsten Bringmann for valuable comments.  M.F. is grateful for funding from the STFC.  P.G. is supported by an ERC grant.
\end{acknowledgments}
\bibliography{arcade_ams_resub3}
\end{document}